\documentclass[12pt]{article}
\usepackage{graphicx}
\usepackage{epsfig}
\usepackage{amssymb}

\usepackage{color}

\def\be{\begin{equation}}
\def\ee{\end{equation}}
\def\ba{\begin{array}{c}}
\def\ea{\end{array}}

\def\ben{$$}
\def\een{$$}

\newcommand{\bea}{\begin{eqnarray}}
\newcommand{\eea}{\end{eqnarray}}

\newcommand{\kkt}{\kt\!\kt}

\newcommand{\pkt}{\!\!\succ\,\,}
\newcommand{\kt}{\rangle}
\begin{document}

\begin{center}

{\Large \bf Paths of unitary access to
exceptional points}

\vspace{0.8cm}

  {\bf Miloslav Znojil}

\vspace{0.2cm}

The Czech Academy of Sciences, Nuclear Physics Institute,

 Hlavn\'{\i} 130,
250 68 \v{R}e\v{z}, Czech Republic

\vspace{0.2cm}

 and

\vspace{0.2cm}

Department of Physics, Faculty of Science, University of Hradec
Kr\'{a}lov\'{e},

Rokitansk\'{e}ho 62, 50003 Hradec Kr\'{a}lov\'{e},
 Czech Republic

\vspace{0.2cm}

{e-mail: znojil@ujf.cas.cz}

\end{center}

%\newpage

\section*{Abstract}

With an innovative ideas of acceptability and usefulness
of the non-Hermitian
representations of Hamiltonians for the description of unitary
quantum systems
(dating back to the Dyson's papers),
the community of quantum physicists was offered a new and
powerful tool for the building of models
of quantum phase transitions.
In this paper the mechanism
of such transitions
is discussed
from the point of view of mathematics.
The emergence of the direct access
to the instant of transition
(i.e., to the Kato's exceptional point)
is attributed to
the underlying split of several
roles played by the traditional single Hilbert space of states ${\cal L}$
into a {\em triplet\,}
(viz., in our notation, spaces ${\cal K}$ and  ${\cal H}$
besides the conventional ${\cal L}$).
Although this explains the
abrupt,
quantum-catastrophic nature of the
change of phase (i.e., the loss of observability)
caused by an infinitesimal change
of parameters,
the explicit description of the unitarity-preserving
corridors of access to the
phenomenologically relevant exceptional points
remained unclear.
In the paper some of the recent results in this direction
are summarized and critically reviewed.

\section*{Keywords}
exceptional points;
quasi-Hermitian quantum theory; perturbations; quantum catastrophes;

\section*{Acknowledgements}

Work supported by the Excellence project P\v{r}F UHK 2020
of the University of Hradec Kr\'{a}lov\'{e}.

\newpage

\section{Introduction.}

In the context of quantum physics the first signs of
appreciation of the phenomenological relevance of
exceptional points \cite{Kato} appeared during the
studies of the so called open quantum systems \cite{Nimrod}.
In these studies
the effective Hamiltonians act in
a model subspace
of the
full Hilbert space and are non-Hermitian.
Thus,
it was not too surprising to reveal that
``the positions of the exceptional points''
vary  ``in the same way as the transition
point of the corresponding phase transition'' \cite{Stellenboschc}.
The possibility of
a generic connection between
exceptional points and phase transitions has been born.

In our present paper we will summarize several aspects of this connection.
In order to narrow the subject
we will only consider the
exceptional-point-related phenomena emerging
in the theory of the closed, stable, unitary quantum systems.

\subsection{Mathematical concept of exceptional point.}

\begin{figure}[h]                    %instead of \begin{figure}[t]
\begin{center}                         %instead of \begin{center}
\epsfig{file=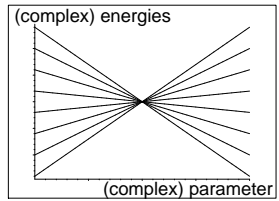,angle=270,width=0.26\textwidth}
\end{center}    % \sidecaption                      %instead of \end{center}
%\vspace{2mm}
\caption{A schematic
sample of the degeneracy of an $N-$plet of energies
at an exceptional point of order $N$ (EPN) with $N=8$ (EP8).
 \label{gloja}}
\end{figure}

In mathematics the exceptional point (EP)
can be defined as the value
of an (in general, complex) parameter $g$
at which
a linear operator (which is, say, non-Hermitian but
analytic in $g$)
loses its diagonalizability. For Hamiltonians,
one of the
possible consequences is schematically depicted in Fig.~\ref{gloja}.
This picture
indicates that
near an EP singularity
there may exist an $N-$plet of
eigenvalues
of the operator
(i.e., typically, bound state energies specified
by a Hamiltonian $H(g)$)
which merge in the limit of $g \to g^{(EP)}$.
Simultaneously, in contrast to the non-EP dynamical scenarios,
the EP or EPN degeneracy also
involves the
eigenvectors \cite{Kato}.

%the Hamiltonian $H(g)$

\subsection{Early applications of exceptional points
in quantum physics of closed systems.}

\begin{figure}[h]                    %instead of \begin{figure}[t]
\begin{center}                         %instead of \begin{center}
\epsfig{file=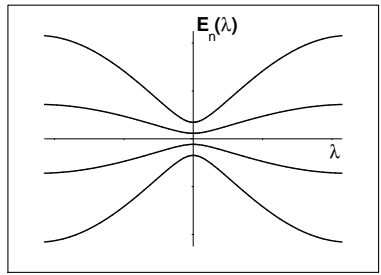,angle=270,width=0.36\textwidth}
\end{center}    % \sidecaption                      %instead of \end{center}
%\vspace{2mm}
\caption{Avoided crossing of four real (i.e., observable)
energy levels (arbitrary units).
 \label{globe}}
\end{figure}

The confluence of eigenvalues
as studied by mathematicians
and sampled by Fig.~\ref{gloja}
did not initially find any
immediate applications in quantum physics of closed systems.
Among the reasons one can find, first of all,
the widespread habit of keeping all of the realistic
phenomenological bound-state
Hamiltonians self-adjoint.
This also required, for pragmatic reasons,
a replacement of the general complex
parameter (say, $g \in \mathbb{C}$ in $H(g)$)
by a
real variable (i.e., by $\lambda \in \mathbb{R}$
in $H(\lambda)$).
A combination of
the two constraints
rendered the
mergers impossible.
Only after an abstract mathematical operation
of analytic continuation of Hamiltonian  $H(\lambda)$
it was possible to reveal,
in several models \cite{BW,Alvarez}, the existence
of the EPs. Naturally, all of
them were manifestly non-real,
${\rm Im}\, \lambda^{(EP)} \neq 0$ \cite{Kato}.
Only an indirect indication of their presence
near a real line of $\lambda$ could have been provided by
the avoided level crossings,
a spectral feature sampled in Fig.~\ref{globe}.

In the quantum unitary-evolution setting
a dramatic change of the situation only came with
the Bender's and Boettcher's pioneering
letter \cite{BB}.
The authors revealed that a suitable weakening
of the property of the
self-adjointness of $H(\lambda)$
could make the EP singularities ``visible'' and
real.
What followed the discovery
(cf. also the later review paper \cite{Carl})
was an enormous increase of interest
of the physics community
in a broad variety of Hamiltonians $H(\lambda)$
possessing the {\em real}
(i.e., in principle, experimentally
accessible) EP singularities with
${\rm Im}\, \lambda^{(EP)} = 0$.
In 2010, the conference
organized by W. D. Heiss in Stellenbosch \cite{Stellenbosch}
was even {\em exclusively} dedicated to the
role of the EPs in multiple branches of physics.

In our present paper we are going to interpret
the EP and EPN degeneracies as sampled by Fig.~\ref{gloja}
in a strict unitary-evolution sense.
This means that
we will only consider the real parameters $\lambda$ lying
in a small vicinity of $\lambda^{(EPN)}$.
Under this assumption we will require
that the whole spectrum of energies remains real
and non-degenerate either
on both sides of $\lambda^{(EPN)}$ (i.e.,
at any not too remote $\lambda\neq \lambda^{(EPN)}$)
or on one side at least (i.e.,
for $\lambda<\lambda^{(EPN)}$
or for $\lambda>\lambda^{(EPN)}$).
We will, naturally, also admit that the value of $\lambda$
parametrizes a smooth curve passing through
a larger, $d-$dimensional
space $\mathbb{R}^d$ of the real parameters determining
a $d-$parametric
Hamiltonian
$H(\lambda)=H[a(\lambda), b(\lambda), \ldots, z(\lambda)]$.

\subsection{Two-parametric example.}

\begin{figure}[h]                    %instead of \begin{figure}[t]
\begin{center}                       %instead of \begin{center}
\epsfig{file=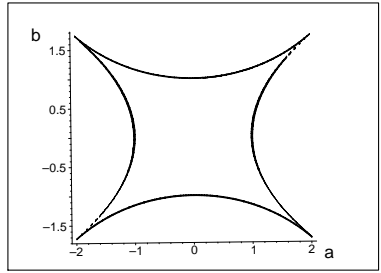,angle=270,width=0.36\textwidth}
\end{center}    % \sidecaption                      %instead of \end{center}
%\vspace{2mm}
\caption{The boundary of domain ${\cal D}^{(physical)}$
for toy-model (\ref{umarnit}) with $d=2$.  \label{fifi}}
\end{figure}

For
illustration let us recall the
two-parametric real-matrix Hamiltonian
of Ref.~\cite{Ruzicka},
  \be
 H(a,b)=\left (\begin {array}{rrrr} -3&b&0&0\\
  -b&-1&a&0 \\
 0&-a&1&b\\
 0&0&-b&3
 \end {array}
 \right )\,.
 \label{umarnit}
  \ee
Its eigenvalues
 \be
 E_{\pm,\pm}(a,b)=\pm \frac{1}{2}\,\sqrt {20-4\,{b}^{2}
 -2\,{a}^{2} \pm 2\,\sqrt {64-64\,{b}^{2}+16\,{a}^{2}+4\,{b}^{2}{a}^{2}+
 {a}^{4}}}\,
  \ee
remain real and non-degenerate inside a two-dimensional
unitarity-supporting
domain ${\cal D}^{(physical)}$
of parameters $a=a(\lambda)$ and $b=b(\lambda)$
which
is displayed in Fig.~\ref{fifi}.

It is worth adding that
once one moves to
the EP-supporting models with
more parameters, $d>2$,
the illustrative
shape of
the $d=2$ domain in Fig.~\ref{fifi}
(viz.,
a deformed square with protruded vertices)
appears to be, in some sense, generic.
For a family of solvable models with $N>4$
such an intuition-based conjecture has been confirmed
in \cite{maximal,tridiagonal}.
A more recent, abstract theoretical explanation of
the hypothesis may be found in \cite{[2],[1]}.
On this background one can expect that
the
most interesting smooth curves
parametrized by $\lambda$ would be those which end
at one of the EPN vertices with maximal $N$ (i.e., with $N=4$
in Fig.~\ref{fifi}).

\subsection{Paradox of stability near exceptional points.}

In a way inspired by the above example one can expect that
the behavior of the closed
quantum system with parameters lying
deeply inside ${\cal D}$ would not be too surprising.
In
such a dynamical regime, small changes of the parameters
leave the spectrum real. The formulation of predictions
can be based on a conventional perturbation theory.

Close to the boundary
$\partial {\cal D}$ the situation is different
and much more interesting.
Indeed, in a small vicinity of this boundary,
a small change of the parameter
seems to be able to cause
an abrupt loss of the observability of the system.
A spontaneous collapse {\it alias\,} quantum phase transition
caused by a small fluctuation of the interaction
seems unavoidable.

Our present paper will be fully devoted to its study.
In fact, a major part of the paper will provide
a concise explanation
that
in the specific context of the closed, unitary quantum systems
the latter, intuitive expectation of instabilities
is incorrect (see section \ref{ksap1} for introduction).
We will clarify
why such an interpretation of dynamics is
incorrect (see section \ref{sekcectyri}),
why a deeper clarification of the point
is important (cf. section  \ref{sekcesest}),
and, finally, what would be a valid
conclusion (section \ref{ystyri}).

\begin{figure}[h]                     %instead of \begin{figure}[t]
\begin{center}                         %instead of \begin{center}
%$\heartsuit$
\epsfig{file=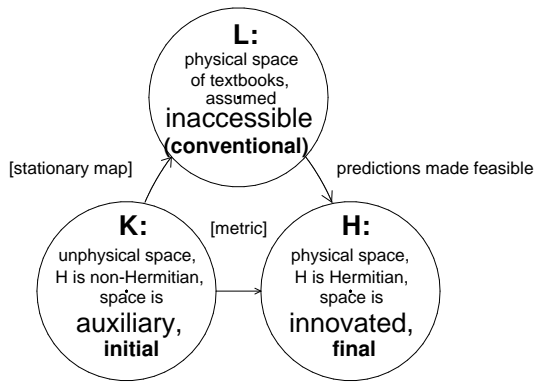,angle=270,width=0.65\textwidth}
\end{center}                         %instead of \end{center}
\vspace{2mm}\caption{{Triplet of Hilbert spaces}
representing a bound state $\psi$ and
connected by
a non-unitary
{map} $\Omega\,$ and by the innovative {\it ad hoc\,} amendment
$I \to \Theta=\Omega^\dagger \Omega \neq I$
of the inner-product
{metric}.}\label{picee3wwww}
\end{figure}

Keeping this purpose in mind, our text
will start by a sketchy presentation of a
(very non-representative) sample
of the current state of applications of the
stationary version of the formalism
represented, schematically, by Fig.~\ref{picee3wwww}. This will be
followed by a (partially critical)
review of some open questions
connected, first of all, with the role of
the Kato's exceptional points in phase transitions.
We will
clarify the role of parameters in
the vicinity of EPs. In this dynamical regime,
a few comments will be also added on the correct
analysis of stability of the non-Hermitian but
unitary quantum systems with respect to
small perturbations.

A concise summary of our present message will finally be
formulated in section \ref{sekcesedm}.

\section{Unitary evolution in Schr\"{o}dinger picture
using non-Hermitian Hamiltonians.\label{ksap1}}

\subsection{Theoretical background}

The above-mentioned paradox of stability near EPs is reminiscent of the
old puzzle of
the stability of atoms in classical physics. In fact, the resolution
of the latter puzzle
belongs among the most remarkable successes which accompanied the birth of
quantum mechanics. The innovation was based on
Schr\"{o}dinger equation
representing bound states
by ket-vector elements of a suitable Hilbert space ${\cal K}$,
 \be
 H\,|\psi_n\kt = E_n\,|\psi_n\kt\,,\ \ \ \ |\psi_n\kt \in {\cal K}
 \,,\ \ \ \ n = 0, 1, \ldots\,.
 \label{se}
 \ee
Subsequently, the incessant
growth of the number of successful phenomenological applications of
quantum theory was
accompanied by the
emergence of various innovative
mathematical subtleties.
One of the ideas of the latter type
(and of a decisive relevance for the present paper)
can be traced back to the papers by
Dyson \cite{Dyson} and Dieudonn\'{e} \cite{Dieudonne}.
Independently, they
introduced the concept of the
$\Theta$-pseudo-Hermitian Hamiltonians.
These operators
(with real spectra) are assumed to remain non-Hermitian
in ${\cal K}$ but restricted by the quasi-Hermiticity
relation
 \be
 H = \Theta^{-1} H^\dagger\Theta \neq H^\dagger\,,
 \ \ \ \ \
 \Theta=\Theta^\dagger>0\,.
 %\Omega^\dagger\Omega \neq I
% \,
 \label{cryher}
 \ee
For details see the text below,
or the older review paper \cite{Geyer},
or its more recent upgrades
\cite{Carl,SIGMA,ali,book,Christodoulides,Carlbook}.

Briefly, the $\Theta$-pseudo-Hermiticity
innovation can be characterized
as a reclassification of the status of the Hilbert space of states
(cf. Fig.~\ref{picee3wwww}).
Indeed, in conventional textbooks
the choice of ${\cal K}$
in Schr\"{o}dinger
Eq.~(\ref{se}) is usually presented as unique.
In the textbook cases of stable, unitarily evolving quantum systems,
in a way observing Stone theorem \cite{Stone},
also the Hamiltonian itself would
necessarily be required self-adjoint
in ${\cal K}$.
After the reclassification, in contrast, the meaning of symbols
${\cal K}$ and $H$ is being changed. Firstly, in the Dyson's spirit one
decides to admit that $H$
can be non-Hermitian in
${\cal K}$. In the light of the Stone theorem this means that
the status of
${\cal K}$ must be changed from ``physical'' to ``unphysical''.
Secondly, in the Dieudonn\'{e}'s spirit,
the postulate of validity of quasi-Hermiticity
relation (\ref{cryher}) enables us to
interpret operator $\Theta$ as a metric \cite{Geyer}.
Thus, we may
amend the
inner product in order to convert
the unphysical Hilbert space ${\cal K}$
into a new, unitarity non-equivalent physical Hilbert space ${\cal H}$,
 \be
 \langle \psi_1|\psi_2\kt_{\cal H}
 =\langle \psi_1|  \Theta|\psi_2\kt_{\cal K}\,.
 \label{innp}
  \ee
Thirdly, the factorization $\Theta= \Omega^\dagger\Omega \neq I$ of the metric
enables us to introduce an operator
 \be
 \mathfrak{h}=\Omega^{-1}\,H\,\Omega
 \label{tri}
 \ee
and to interpret is as a hypothetical alternative
isospectral Hamiltonian in another, alternative physical
Hilbert space ${\cal L}$ which is, by the assumption
which dates back to Dyson \cite{Dyson,Geyer}, self-adjoint but
constructively as well as technically inaccessible.

\subsection{Notation conventions}

Further details characterizing such
an apparently redundant representation
of a single state $\psi$
will be recalled and summarized below.
Now, let us only
point out that  the Dyson's and Dieudonn\'{e}'s
reformulation of the postulates of quantum theory was
deeply motivated. It is only necessary to
accept and appreciate
{\em both\,}
the Dyson's activity {\em and\,} the
Diedonn\'{e}'s scepticism.
Indeed, Dyson discovered and used, constructively,
several positive and truly innovative aspects of
the use of quasi-Hermiticity in physics and phenomenology.
At the same time, the Dieudonn\'{e}'s
well founded critical analysis
of the ``hidden dangers'' behind the quasi-Hermiticity is
still a nontrivial
and exciting subject for mathematicians \cite{Trefethen,ATbook}).

In the context of physics the latter ``hidden dangers'' were,
fortunately, cleverly
circumvented (cf. review \cite{Geyer}). Some of
the corresponding technical and mathematical recommendations
will be recollected below. Immediately,
let us only mention that the amendment
of mathematics led to an ultimate compact and explicit
three-Hilbert-space (3HS) formulation of
the most general non-stationary version of quasi-Hermitian quantum
mechanics as first proposed in \cite{timedep} and as
subsequently reviewed in \cite{SIGMA}.

In what follows, we will employ the most compact notation
as introduced in our latter two
papers. The reason is twofold. Firstly,
along the lines indicated in \cite{SIGMA}, the choice of
such a notation
will simplify the separation of our present perception of physics
from its alternatives which often share the mathematical terminology
while not sharing
the phenomenological scope.
Secondly, the emphasis put on
notation will enable us to
review the field of
our present interest in a sufficiently compact and concise manner,
avoiding potential misunderstandings caused by the
variabiity of the notation used in the literature
(cf. Table~\ref{dowe}).

\begin{table}[h]
\caption{Sample of confusing differences in notation conventions.}
 \label{dowe} \vspace{.4cm}
%\centering \label{tataj}
\centering
\begin{tabular}{||c||c|c|c|c||}
    \hline \hline %&&&&&\\
    {\rm concept  }
    &
    \multicolumn{4}{|c||}{symbol}\\
% &
\hline
\hline
%  {\rm observable Hamiltonian} &  $H$  & $\widetilde{H}$&$H$ &$H$\\
 %$=\Omega^\dagger\Omega$, the
   {\rm Hilbert space metric}
    &  $\eta_+$  & $\rho$&$\widetilde{T}$&
 $\Theta$\\
  {\rm Dyson's map}
    &  $\rho$  & $\eta$&$S$ &
 $\Omega$ \\
   {\rm state vector}
    &  $|\psi\kt$  & $\Psi$ &$|\Psi\kt$&
 $|\psi\kt$\\
   {\rm dual state vector}
    &  $|\phi\kt$  &
    $\rho\Psi$ &$\widetilde{T}|\Psi\kt$&$|\psi\kkt$\\
% &  {\rm the generator of kets} &  $-$  & $H$&$-$ &$G$\\
% %&  {\rm textbook Hamiltonian}&  $h$  & $h$ &$-$ &$\mathfrak{h}$\\
% &  {\rm Coriolis Hamiltonian}&  $-$  & unabbr.  &$-$& $\Sigma$\\ &&&&&\\
\hline \hline
{\rm reference}&
    \cite{ali} &
    \cite{Fring15b} &
    \cite{Geyer}&
   here
    \\
 \hline
 \hline
\end{tabular}
\end{table}

\subsection{The concept of hidden Hermiticity.\label{sekcedva}}

%Implementations of the idea of preconditioning

\subsubsection{Motivation.\label{susscectyri}}

In an incomplete sample of ambitions of the 3HS reformulation of
quantum theory
let us mention, first of all,
the Dyson's description of
correlations in many-body systems \cite{Dyson}
inspired by
numerical mathematics
(where one would speak simply about a
``preconditioning'' of the Hamiltonian).
Secondly,
in combination with the assumption of ${\cal PT}-$symmetry \cite{Carlbook}
the 3HS approach
(complemented by the mathematical Krein-space methods \cite{Langer,AKbook})
opened new horizons in our understanding of the
first-quantized relativistic
Klein-Gordon equation \cite{aliKG,jaKG}.
Thirdly,
a transfer of the underlying ``hidden Hermiticity'' ideas to
relativistic quantum field theory \cite{BM}
and/or to the studies of supersymmetry \cite{Ioffe}
inspired a number of methodical studies
of various elementary toy models \cite{BB,BG,Mateo}.
Last but not least it is worth mentioning that
the applications of the 3HS formalism
even touched the field of canonical quantum gravity
based on the use of Wheeler-DeWitt equation \cite{WDW}.

\subsubsection{Disambiguation.\label{sekcetri}}

The solution  $\Theta=\Theta(H)$ of Eq.~(\ref{cryher}) does not exist
whenever the spectrum of $H$ ceases to be real.
This means that only certain
parameters in non-Hermitian $H(\lambda)$
remain unitarity-compatible and ``admissible'',
$\lambda \in {\cal D}$.
In the admissible cases,
in a way explained in \cite{SIGMA},
there exists a mapping $\Omega$ which
realizes an equivalence of predictions
made in ${\cal H}$ with those made
in a third, hypothetical and, by our assumption,
practically inaccessible
Hilbert space ${\cal L}$. The latter space is
precisely the space of states used in
conventional textbooks. In the present 3HS context
(depicted in Figure \ref{picee3wwww}),
its role is purely formal because in this space,
operator (\ref{tri})
representing the Hamiltonian
and formally self-adjoint in ${\cal L}$
is, by assumption,
too complicated
to be useful or tractable (for example,
it may happen to be a highly non-local
pseudo-differential operator \cite{ali}).

In the
literature devoted to applications of unitary quantum theory
the authors
working in the
3HS version of Schr\"{o}dinger picture
do not always sufficiently clearly
emphasize the
Hermiticity
of the physical Hamiltonian
in the physical Hilbert space ${\cal H}={\cal H}^{(Standard)}$
(say, by writing $H=H^\ddagger$ \cite{SIGMA}).
Another potential source of confusion lies in
the widespread habit (or rather in the abuse of language)
of using
shorthand phrases (like ``non-Hermitian Hamiltonians'')
or
shorthand formulae (like $H\neq H^\dagger$)
without adding that one just temporarily dwells in an irrelevant,
auxiliary, unphysical Hilbert space ${\cal K}$.
The resulting, fairly high probability of misunderstandings is further enhanced
by the diversity of conventions as sampled in Table~\ref{dowe}.

\section{Constructive aspects of the
triple Hilbert space formalism.\label{sekcectyri}}

\subsection{Metric and its ambiguity.\label{ssekcctyri}}

Two alternative model-building strategies based on the
``generalized Hermiticity'' (\ref{cryher}) have been used
in applications. In the first one
one chooses $\mathfrak{h}=\mathfrak{h}^\dagger$
and $\Omega$ and reconstructs  $H$ and  $\Theta$.
In fact, the use of
such a strategy
remained restricted just to
nuclear physics of heavy nuclei
in practice \cite{Geyer}.
At present, almost exclusively \cite{ali},
one picks up
the Hamiltonian (i.e.,
a ``trial and error'' operator $H$
which is non-Hermitian in ${\cal K}$)
and reconstructs, via Eq. (\ref{cryher}), the
(necessarily, nontrivial)
metric $\Theta>0$
(i.e., the correct physical
Hilbert space of states denoted,
here, by dedicated symbol ${\cal H}$).
The approach
based on the reconstruction of metric
now forms the mainstream in research.
The false but friendly space ${\cal K}$ and a
non-Hermitian {Hamiltonian}
$H$ are both assumed to be given in advance while
a suitable Hermitizing inner-product {metric} must be reconstructed,
in principle at least.
The Hermiticity of any other
observable $\Lambda$ in ${\cal H}$
must
also be guaranteed.
In the auxiliary space ${\cal K}$
this requirement has the form $\Lambda^\dagger\,\Theta=\Theta\,\Lambda$.

In an
elementary illustration the Wheeler-DeWitt-like equation
 \be
 H=H^{(WDW)}(\tau)=
 \left[ \begin {array}{cc} 0&\exp 2 \tau\\
 \noalign{\medskip}1&0\end {array} \right]
 \neq H^\dagger\,\ \ \ \ \ {\rm in}\ \ \
 {\cal K}=\mathbb{R}^2
 \label{primo}
 \ee
yields the two real closed-form eigenvalues $E= E_\pm=\pm \exp \tau$
so that it
can serve as a sample of the Dyson-Dieudonn\'{e} definition
of
quasi-Hermiticity (\ref{cryher}).
A decisive advantage of
the use of
such a highly schematic
one-parametric
two-by-two real-matrix example
is that one can easily solve Eq.~(\ref{cryher})
and construct
{\em all\,} of the eligible
physical
inner-product-metric operators
 \be
 \Theta=\Theta^{(WDW)}(\tau,\beta)=
 \left[ \begin {array}{cc} \exp (-\tau)&\beta\\
 \noalign{\medskip}\beta&\exp \tau\end {array} \right]
 =\Theta^\dagger\,,
 \ \ \ \ \   |\beta|<1 \,.
 \label{secuf}
 \ee
These solutions
form a complete set of
candidates
for the (Hermitian and positive definite) eligible
metric \cite{Geyer}.
In this example
one notices that
parameter $\beta$ is an independent variable.
This observation is, indeed, compatible with the well known fact that
the assignment
$\Theta=\Theta(H)$ of the metric to a preselected Hamiltonian
is not unique \cite{Geyer,ali,Lotoreichik}.

%\section{Methodical considerations and schematic models.\label{pasekcesest}}

\subsection{False instabilities and open systems in disguise \label{whystyri}}

In the
literature devoted to applications
the authors interested in non-Hermiticity often
do not sufficiently clearly
separate the quantum and non-quantum
theories. Here, we are not going to deal with the latter
branch of physics. Nevertheless, even within the range of
quantum mechanics
the authors often intermingle the results concerning the
open and closed quantum systems.
Here, almost no attention will be paid to the former
family of models, either.
An exception should be made
in connection with the papers
dealing with certain
non-Hermitian but
${\cal PT}-$symmetric
quantum systems where, typically, the authors claim
that
``complex eigenvalues may appear very far from
the unperturbed real ones
despite the norm of the perturbation is
arbitrarily small'' \cite{Viola}.

As long as the latter claims
(of a top {\em mathematical\,} quality)
are accompanied
by certain fairly vague quantum-theoretical considerations
(which could certainly prove misleading),
we feel forced to point out that
recently, the study of the parametric
domains of unitarity near EPs
\cite{[3]} clarified the point
(cf. also the less formal explanation in
\cite{Ruzicka}).
The essence of the misunderstanding
can be traced back to the fact that
the loss of stability
was deduced,
in \cite{Viola}, from the properties of the
pseudospectrum \cite{Trefethen}.
Unfortunately,
the construction was only performed using the
trivial form of the inner-product metric defining
just the
manifestly unphysical Hilbert space ${\cal K}$ where $\Theta=I$.
For this reason the mathematical results
about pseudospectra in ${\cal K}$
make sense in, and only in, the open quantum systems.
In these systems the predicted
instabilities really do occur because
the space ${\cal K}$ itself
still keeps there the status of the physical
space.

We may summarize that in the 3HS models of closed systems
the Hamiltonians are in fact self-adjoint in ${\cal H}$.
This means that the evaluation of their pseudospectra would
{\em necessarily\,} require the work with norms which would
be expressed
in terms of the physical
metric $\Theta$.
Thus, once
the existence of such a metric is guaranteed
(which is, naturally, a nontrivial task!),
the
proofs of stability
based on the pseudospectra will apply.

We should also add that the smallness of
perturbations
is a concept which crucially depends on the metric $\Theta$
defining the physical Hilbert space ${\cal H}$.
From this point of view
it is obvious that
as long as the metric itself becomes necessarily strongly
anisotropic in the vicinity of EPs \cite{Lotoreichik}, also
some of
the perturbations which might look small in  ${\cal K}$
become, in such a regime, large in ${\cal H}$,
and {\it vice versa} \cite{[5]}.

\subsection{EP (hyper)surfaces and their geometry.}

For the lovers of closed formulae
the existence
as well as the geometry of access to EPs
was made very explicit in paper
\cite{[1]}.
An advertisement of
the contents of this paper
can be brief:
a list of transmutations
is given there between various versions of
a special Bose-Hubbard (BH) system
(represented by certain complex finite matrices)
and of a discrete and truncated
anharmonic oscillator (AO). It is sufficient to
recall here just the ultimate message of the paper:
at an EP singularity of order $N$ it is possible to match,
via a phase transition, many entirely
different quantum systems.
Represented in
their respective Hilbert spaces ${\cal K}$ and
sharing just their dimension $N<\infty$.

In
\cite{[1]} the idea is illustrated via its
several closed-form realizations.
Incidentally, all of these models happened to be
unitary in a domain ${\cal D}$
of a shape resembling a (hyper)cube with
protruded vertices.
In a broader perspective one can say that
by definition \cite{Kato}, the latter vertices
are precisely the EP extremes of our present interest.
In this light,
our present paper could be briefly characterized
as a study of the geometry
of the generic unitarity-supporting domains of parameters,
with particular emphasis on understanding
of the sharply spiked shapes of their surfaces $\partial {\cal D}$
in a small vicinity of their EP vertices and edges.
Indeed, we found such phenomenologically relevant
features of the geometry
mathematically remarkable and worth a dedicated study.

\section{Real-world models and predictions.\label{sekcesest}}
%\subsection{The role of exceptional points.\label{sekcepet}}

\subsection{Mathematics: Amended inner products
and exceptional points.\label{scectyri}}

The main purpose of the introductory
recollection of the 3HS formalism
was to prepare a turn of attention to
the key role played, in the 3HS applications,
by the concept of exceptional points (EPs).
Although their original rigorous definition may be already
found in the old
Kato's monograph on perturbation theory \cite{Kato},
their usefulness for quantum physics of unitary systems
only started emerging after Bender with Boettcher
pointed out, in their pioneering letter  \cite{BB},
that the EPs
(also known as Bender-Wu singularities \cite{BW,Alvarez})
could also acquire an immediate phenomenological
interpretation of the points of quantum phase transition.
Alternatively, their properties appeared relevant in the more speculative
contexts of Calogero models and/or of supersymmetry
\cite{[9]}. % Znojil, "Supersymmetry and exceptional points."

From all of the similar 3HS-applicability points of view
it is necessary to start the model-building processes
from a preselected candidate for the Hamiltonian which
is parameter-dependent, $H=H(\lambda)$.
Moreover, it must be non-Hermitian in
the auxiliary Hilbert space ${\cal K}$ and,, at the same time,
properly Hermitian and self-adjoint
in an ``amended'' Hilbert space of states ${\cal H}$.
Now, the key point is that
in the light of assumption (\ref{cryher}),
the latter space can be represented
via a mere amendment (\ref{innp}) of the inner product in ${\cal K}$.
In other words, {\em any\,}
solution $\Theta=\Theta(H)$ of Eq.~(\ref{cryher})
{\em defines\,} the necessary physical space ${\cal H} ={\cal H}(H)$.

In opposite direction,
{\em many\,} of the
eligible and Hamiltonian-dependent metrics and spaces
may and will {\em cease\,} to exist
{\em before\,}
the variable, path-specifying
parameter $\lambda$ reaches the ultimate EP value
$\lambda^{(EP)} \in \partial {\cal D}$.
For the
parameters lying inside the physical domain ${\cal D}$,
the Hamiltonian
must still be assigned
such a specific metric $\Theta$ and space ${\cal H}$
which would exist {\em up to the required limit\,}
of $\lambda \to \lambda^{(EPN)}$.
In this sense, for any preselected quasi-Hermitian
quantum system,
our knowledge and specification of the boundary $\partial {\cal D}$ near EPNs
are of an uttermost importance.

\subsection{Realistic many-body systems.}

In the latter setting we should return,
once more, to
Fig.~\ref{fifi} illustrating
the sharply spiked, fragile,
parameter-fine-tuning
nature of the shape of the sample domain near its EPN extremes.
Due to their potential phase-transition interpretation,
these extremes seem to be the best targets of
a realistic experimental search.

\subsubsection{Realistic systems inclined to support
an approximate decomposition into clusters.}

The manifestly non-unitary mapping $\Omega$
as mentioned in Fig.~\ref{picee3wwww} connects the
ket-vector elements of two non-equivalent Hilbert spaces:
In the notation of Ref.~\cite{SIGMA} we have
  \be
  |\psi\pkt = \Omega\,|\psi\kt\,,\ \ \ \
  |\psi\pkt \in {\cal L}\,,\ \ \ \
  |\psi\kt \in {\cal K}\,.
  \label{fakto}
  \ee
Recently it has been revealed that
precisely the same mapping
(attributed to Dyson \cite{Geyer})
also forms a mathematical
background of the so called
coupled cluster method (CCM, \cite{Cizek}).
In fact, the
implementation aspects of the
latter, CCM interpretation of formula (\ref{fakto})
were already used in calculations
and tested, say, in quantum chemistry.
What was particularly successful are the
variational (or, more precisely, bi-variational)
realizations of the CCM philosophy, with
emphasis put upon the construction of ground states, and with a
well-founded
preference of mappings (\ref{fakto}) in the
exponential form $\Omega = \exp S$
where $S$ is represented in a suitable operator basis.

The latter, apparently purely technical restriction
seems to be responsible for the
success of the method
which is currently
``one of the most versatile and most accurate of
all available formulations of quantum many-body
theory''
\cite{[10]}.
In paper
\cite{[10]},
extensive 3HS-CCM parallels
have been found.
The respective strengths and weaknesses of the two approaches
look mutually complementary.
Currently \cite{ZB}, their further analysis is being concentrated upon the
strengths. One may expect that the
consequent, mathematically consistent 3HS quantum theory
might enhance the range of applicability of the more pragmatic but
very
precise
CCM ground-state constructions.
Along these lines, in particular,  the new theoretical predictions
may be expected to concern the EP-related
many-body quantum
phase transitions
which could be also, in parallel, experimentally detected.

\subsubsection{Bose-Hubbard model and its open- and
closed-system interpretations.}

The Bose-Hubbard Hamiltonian
 \be
  \label{Ham1}
  H = \varepsilon\left(a_1^{\dagger}a_1 - a_2^{\dagger}a_2\right) +
  v\left(a_1^{\dagger}a_2 + a_2^{\dagger}a_1\right) + \frac{c}{2}
  \left( a_1^{\dagger}a_1 - a_2^{\dagger}a_2\right)^2
 \ee
of Graefe et al \cite{Uwe} has been developed to
{describe} an $(N-1)$-particle Bose-Einstein condensate in a double
well potential containing a sink and a source of equal strengths.
Besides the usual annihilation and
creation operators the definition contains
the purely imaginary on-site energy difference
$2\varepsilon = 2{\rm i} \gamma$.
In the
{fixed$-N$} representation
the Hamiltonian is a {matrix}: At
$N=6$ we have, for example,
%$
%H(c)= H(0) + c\,V\,=
%$
\be
\label{example1}
H^{(6)}(\gamma,c,v)= \left(
\begin{array}{cccccc}
 -5 {\rm i} \gamma+ \frac{25}{2} c & \sqrt{5} v & 0 & 0 & 0 & 0\\
 \sqrt{5} v & -3{\rm i}\gamma+\frac{9}{2}c & 2\sqrt{2} v & 0 & 0 & 0\\
 0 & 2\sqrt{2}v & -{\rm i}\gamma+\frac{1}{2}c & 3v & 0 & 0\\
 0 & 0 & 3v & {\rm i}\gamma+\frac{1}{2}c & 2\sqrt{2}v & 0\\
 0 & 0 & 0 & 2\sqrt{2}v & 3{\rm i}\gamma+\frac{9}{2}c & \sqrt{5}v\\
 0 & 0 & 0 & 0 & \sqrt{5}v & 5{\rm i}\gamma+\frac{25}{2}c
\end{array}
\right)\,.
\ee
Once we fix the inessential
single particle
tunneling constant $v=1$
and once we localize the EPN singularity at $\gamma=1$ and at
the vanishing
strength of the interaction between particles $c = 0$,
we reveal, at any $N$,
that
after an {\it  arbitrarily small\,} {$c \neq 0$} perturbation,
the spectrum {\em abruptly\,} ceases to be real (see {\it loc. cit.}).
This means that the metric $\Theta$
and space ${\cal H}$
cease to exist, either.
The perturbed system admits, exclusively, the
non-unitary, open-system interpretation in ${\cal K}$.

In our present framework restricted
to closed systems, only the parameters contained inside
the suitable physical domain
${\cal D} = \{\gamma,
c\,|\, \gamma \in (-1,1)\,,c \in (c_{\min}(\gamma),c_{\max}(\gamma))\}$
(with the shape resembling, locally, Fig.~\ref{fifi} near its spikes)
would be compatible with the reality of the energies.
Interested readers may find an extensive
study and detailed constructive description of the shape of
such a unitarity compatible domain in
our rather lengthy recent paper~\cite{[13]}.

\subsection{Generalized Bose Hubbard models}

Up to now we paid attention to the models
(sampled by the Bose Hubbard Hamiltonian (\ref{Ham1})) with
the EPN singularities possessing the trivial
geometric multiplicity $K=1$  \cite{Kato}.
Interested readers may find,
in paper  \cite{[6]},
an introduction into a more general
category of the EPs
characterized by a clustered,
$K-$centered degeneracy of the wave functions
with $K>1$.
In these cases the EP-related quantum
catastrophes (i.e., the generalized  ${\cal PT}-$symmetry
breakdowns) appeared to be of the form
of confluence of several independent EPs with $K=1$.
The paper illustrated the
advanced mathematics of the degeneracy of degeneracies
via low-dimensional matrix
models. The emergence of unusual horizons
found its mathematical formulation in the
language of geometry of Riemann surfaces, accompanied by the
phenomenological predictions of certain anomalous phase transitions.

A model-independent analysis of these
anomalies in the dynamical EP-unfolding scenarios
was based, in subsequent paper
\cite{[12]}, on their
parametrization by the matrix elements of
admissible (i.e., properly scaled and unitarity-compatible) perturbations.
A consistency of algebra with
the EP-related deformations of the Hilbert-space geometry has been confirmed.
The new degenerate perturbation techniques
were developed and their implementation has been found feasible.
Via a class of
schematic models, a constructive analysis of the vicinity of the simplest
nontrivial EPN with $K=2$ was performed.

An implementation of the schematic recipe to the Bose-Hubbard-type
generalized models
may finally be found described in \cite{[13]}. It was shown that
there always exists a non-empty unitarity
domain ${\cal D}$ comprising a multiplet of
variable matrix elements of the admissible perturbations for which the
spectrum is all real and non-degenerate.
The intuitive expectations were confirmed:
the physical parametric domains near EPs were found sharply spiked.

A richer structure was revealed to characterize the admissibility
of the perturbations.
Two categories of the models were considered. In the first one
the number of bosons
was assumed conserved
(leading to the matrix Hamiltonians of the form (\ref{example1})).
The alternative assumption of
the particle-number non-conservation
led to the realistic $K>1$ scenarios
in which the spectra also remain
real and non-degenerate.
The quantum evolution controlled by the
Hamiltonians of larger (or even infinite)
dimensions still remains unitary.

In all of these cases,
in spite of a rapid increase of the complexity of the formulae
with the number of particles,
the existence as well as
a sharply spiked structure of ${\cal D}$ near EPN
has again been reconfirmed.
The first steps of the explicit constructive analysis
of the structure of ${\cal D}$
were performed
in
the simplest case with $N=5$
where the access to EP5
appeared mediated by
eight independently variable parameters.

\subsection{Further phenomenological challenges.}

The early abstract words of warning against the
deceptive nature
of the concept of quasi-Hermiticity
\cite{Dieudonne,Trefethen} were
recently reconfirmed by the authors of paper \cite{Siegl}.
After a detailed analysis of the popular non-Hermitian but
${\cal PT}-$symmetric imaginary
cubic anharmonic oscillator
these authors came to the conclusion that
such a {\it ``fons et origo''} of the theory
can be characterized by the singular behavior attributed to
an ``intrinsic'' EP.
Such a discovery contributed to
the motivation of our present study since it
enhanced the importance of the knowledge of the
behavior of the 3HS models at parameters lying close to
their EP limits.

Another, independent source of interest in the study and
explicit description of the domains ${\cal D}$ of
the unitarity-compatible ``admissible'' parameters
in the close vicinity of
EPs may be seen in the frequently experimentally observed
phenomenon of the avoided level crossings.
In a way sampled by Figure \ref{fifi}
this phenomenon occurs even in the spectra of finite-dimensional
Hermitian matrices.

The related, highly desirable
analytic continuation of the spectra towards their EP
degeneracies is by far not an easy task.
The task is intimately connected with
the 3HS-inspired turn of attention
to the description of quantum dynamics using non-Hermitian Hamiltonians.
This opens multiple technical questions.
One of them is that
after one perturbs a quasi-Hermitian Hamiltonian
or even only its parameter, $H(\lambda) \to H(\lambda')$,
one immediately encounters the re-emergence of the well known ambiguity of the
Hilbert-space inner product in Eq.~(\ref{innp}) \cite{ali,Lotoreichik,arabky}.
As a byproduct of this observation there
appeared a need of
a deep and thorough reformulation of perturbation theory itself
\cite{[5]},
with nontrivial consequences
concerning, in particular, the systems lying close to the
boundary $\partial {\cal D}$.

%********** Figure 1 zde
%plot(exp(-2*x)-2*exp(-x),x=-1.02..5,tickmarks=[4,4],axes=normal);
\begin{figure}[b]                    %instead of \begin{figure}[t]
\begin{center}                         %instead of \begin{center}
\epsfig{file=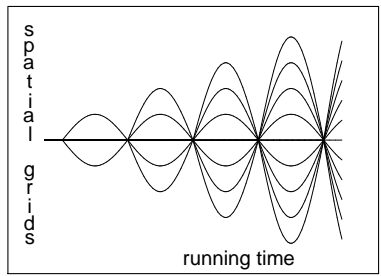,angle=270,width=0.36\textwidth}
\end{center}    % \sidecaption                      %instead of \end{center}
\caption{Schematic 3HS picture of the Universe
evolving through a sequence of
Eons separated by EPs.
Sampled
by ``breathing''
one-dimensional  $N-$point grids with $N_1=2$, $N_2=4$, etc.
 \label{fitva}
 }
\end{figure}

Besides the technical open questions there also exists a
number of the
strong parallel challenges
emerging in the context of quantum phenomenology.
Their truly prominent samples
emerged in the context of quantum cosmology
and, in particular, in
the attempted
descriptions of the evolution of the Universe
shortly after its initial
Big Bang singularity.
The key point is that the classical-theory-supported existence
of Big Bang
seems to
contradict the conventional
quantum-theoretical paradigm of Hermitian theory.
By the latter theory the Big-Bang-type
phase transitions cannot exist,
being ``smeared'' and reduced to
the mere avoided crossing behavior
of the spatial coordinates
called Big Bounce of the Universe \cite{Ashtekar}.

A disentanglement of the puzzle could be, in principle, offered
by the 3HS models in which the Big Bang would
correspond to a real EP-related spectral singularity of
a suitable non-Hermitian operator (cf., e.g., \cite{[7]}).
Such a hypothesis would admit even a highly speculative
``evolutionary cosmology'' pattern of Fig.~\ref{fitva} in which
a sequence of penrosian Eons
separated by the Big-Crunch/Big-Bang singularities would
render the structure of the ``younger'' Universes
richer and more sophisticated.

\section{Exceptional-point-mediated quantum phase transitions.\label{ystyri}}

\subsection{EPs as quantum crossroads.\label{sekcesestapul}}

In paper \cite{[1]} we emphasized that
a classification of
passages of closed quantum systems through
their EP singularities
could be perceived as a quantum
analogue of the classical catastrophe theory \cite{Zeeman}.
In this context let us only add that the
EP-mediated phase transitions
could acquire the form of a quantum
process of
bifurcation,
 \ben
 \ba
 \begin{array}{|c|}
  \hline
 \vspace{-0.35cm}\\
      % {\rm the\ time\ ,\\
  % {\rm initial}\\
         {\rm initial\ phase,}\ t<0, \\
    {\rm  Hamiltonian\ } H^{(-)}(t)\\
 % {\rm {for\ a}\ clearer \ {\rm physical}\ interpretation} \\
%  {\rm  one \ may\ reconstruct\ also\ the\  conventional }\\
% {\rm  Hamiltonian }
% \
% \mathfrak{h}
% =\Omega H\Omega^{-1}=
% \mathfrak{h}^\dagger\ {\rm and\ the}\\
%  {\rm  \fbox{\rm {{\bf textbook} Hilbert space  ${\cal L}$}}.}\\
 \hline
 \ea
 %\\ \ \ \ \ \ \ \
%\stackrel{{ \bf the\ Dyson's\ correspondence} }{}
%  \ \nwarrow\!\!\!\searrow\ \  \  \ \ \ \ \ \
 %\ \ \ \ \ \ \ \
% \ \ \ \ \ \ \ \
% \ \ \ \ \ \ \ \
% \ \ \ \ \ \ \ \
 % \ \ \ \ \
\\
  \Downarrow \,\\
    % \downarrow\ ``on\ the\  road''
    \stackrel
 %{\large \bf  (EP\ limit)}
 { \bf process \ of\ degeneracy}{\rm }
 \\
  \Downarrow \,\\
  %\vspace{-1cm}
  %\ba
   %\vspace{-0.3cm}\\
    \begin{array}{|c|}
 \hline
 \vspace{-0.35cm}\\
   %{\rm option\ A\!:}\\
   % {\rm the\ real-matrix}\\
%   {\rm anharmonic - oscillator}\\
 t>0\ {\rm \ branch\ A,}\\
    {\rm \ Hamiltonian\ } H^{(+)}_{(A)}(t)\\
%  {\rm metric\ } \\
%  \Theta=\Omega^\dagger\Omega \ \ ({\rm s.\ t. }\
%   H^\dagger \Theta=\Theta
%  H),\ i.e.,\\
%  {\rm \fbox{\rm {{\bf physical} Hilbert space  ${\cal H}$}}}\
%   {\rm in \ which }\\
%   {\rm the\ evolution\ becomes\ \bf unitary;} \\
%   %
%  {\rm {is\ made\   selfadjoint\ {\rm via }}}
% \\
  \hline
 \ea
 \stackrel{ {\bf   option\ A}  }{ \Longleftarrow }
 \begin{array}{|c|}
 \hline
 \vspace{-0.35cm}\\
  % {\rm the\ {EP}\  instant,}\\
    %{\rm indeterminacy\!:}\\
      {\rm the\ EP\ ``crossroad",}\\
  {\rm indeterminacy\ at}\ t=0 \\
%  {\rm metric\ } \\
%  \Theta=\Omega^\dagger\Omega \ \ ({\rm s.\ t. }\
%   H^\dagger \Theta=\Theta
%  H),\ i.e.,\\
%  {\rm \fbox{\rm {{\bf physical} Hilbert space  ${\cal H}$}}}\
%   {\rm in \ which }\\
%   {\rm the\ evolution\ becomes\ \bf unitary;} \\
   %
%  {\rm {is\ made\   selfadjoint\ {\rm via }}}
% \\
 \hline
 \ea
 \stackrel{ {\bf  option\ B}  }{ \Longrightarrow }
 \begin{array}{|c|}
 \hline
 \vspace{-0.35cm}\\
   % t>0\ {\rm option\ B\!:}\\
    %{\rm  complex-symmetric}\\
   %{\rm Bose-Hubbard}\\
 t>0\ {\rm \ branch\ B,}\\
    {\rm \ Hamiltonian\ } H^{(+)}_{(B)}(t)\\
  % {\rm one\ picks\ up\ an\ {unphysical} \ but}\\
%  {\rm  \fbox{\rm {{\bf user-friendly} Hilbert space  ${\cal K}$}}\ and\ a}\\
%     {\rm \bf non\!-\!Hermitian\  } H\
%     {\rm with\ real\ spectrum,}\\
%    {\rm {\it i.e.,}\  a\ \bf {\it bona\ fide} \  Hamiltonian};
% \\
%  %\vspace{-0.3cm}\\
 \hline
 \ea \
  \\
 \ea
 %\label{humandiag}
 \een
Thus, in principle, the future extensions of our present models
might even incorporate a multiverse-resembling branching of
evolutions at $t=0$.

Marginally, let us add that in such a branched-evolution setting
one could find applications
even for some results on non-unitary,
spectral-reality-violating evolutions.
An illustration may be found in papers
(sampled by \cite{[8]})
where just
the search for the EP degeneracies
has been performed without any efforts of
guaranteeing the reality of the spectrum.

\subsection{Perturbation theory near EPs
using nonstandard unperturbed Hamiltonians.}

At the above-mentioned ``cross-road'' EP instant $t=0$ the
Hamiltonian ceases to be diagonalizable.
This means that such an instant can be perceived as
a genuine quantum analogue of the
classical Thom's bifurcation singularity {\it alias\,}
catastrophe \cite{Zeeman}.
The distinguishing
feature of the phenomenon
in its quantum form is that it is
``instantaneously'' incompatible with
the postulates of quantum theory.
Fortunately,
the theory returns in full force at
any, arbitrarily small time before or after the catastrophe.

In \cite{[1]}, several explicit and strictly algebraic,
solvable-model illustrations
of such a passage through the EPN singularity may be found
described in full detail. Alternatively, the
phenomenon can be also described in a model-independent
manner. Indeed, a return to the diagonalizability
can be characterized as a perturbation of the
non-diagonalizable $t=0$
Hamiltonian
$H_{(EP)}$.
Thus,
any multiplet of states $|\vec{\psi}(t)\kt$
can be constructed, before or after $t=0$, as the
solution of a properly perturbed Schr\"{o}dinger equation
 \be
 (H_{(EP)}+\lambda\,W)
 |\vec{\psi}\kt = \epsilon \, |\vec{\psi}\kt\,.
 \label{purp}
 \ee
One has to keep in mind that
the unperturbed Hamiltonian itself is
an anomalous operator, the
conventional diagonalization
(or, more generally, spectral representation) of which
does not exist.
Once we have to consider here just its finite-dimensional
matrix forms,
the constructive approach to Eq.~(\ref{purp})
can be based on the
evaluation
of the so called transition matrices $Q_{(EP)}$,
defined as
solutions of the
Schr\"{o}dinger-like linear algebraic equation
 $$
 H_{(EP)}Q_{(EP)} = Q_{(EP)}J_{(JB)}(E_{(EP)})\,.
 $$
The symbol
$J_{(JB)}(E_{(EP)})$
denotes here the canonical representation
of $H_{(EP)}$. Once we decide to choose it in the
most common Jordan-matrix form,
the related transition matrices $Q_{(EP)}$
can be reinterpreted as an analogue of the unperturbed basis.
In this basis,
the  perturbed Schr\"{o}dinger equation (\ref{purp})
acquires the canonical form
 \be
 [J_{(JB)}(E_{(EP)})+\lambda\,V]
 |\vec{\phi}\kt = \epsilon \, |\vec{\phi}\kt\,.
 \label{purpura}
 \ee
Interested readers are recommended to consult Refs.~\cite{[3]}
and ~\cite{[13]} for the further details of the solution
of such an equation.

For our present purposes the essence of the latter technicalities
may be explained using the elementary
unperturbed real-matrix Hamiltonians
of Ref.~\cite{[1]},
 \ben
 H^{(2)}_{(EP)} = \left [\begin {array}{cc} 1&1\\{}-1&-1
 \end {array}\right
 ]\,,
 \ \ \ \ \ \
 H^{(3)}_{(EP)} = \left [\begin {array}{ccc} 2&\sqrt{2}&0\\{}-\sqrt{2}&0
 &\sqrt{2}\\{}0&-\sqrt{2}&-2\end {array}\right ],\ \ldots\,.
 \een
For
this series of examples all of
the transition matrices are non-unitary but known in closed form.
At $N=3$ one gets
 \ben
  Q^{(3)}_{(EP)} = \left[ \begin {array}{ccc} 2&2&1
 \\\noalign{\medskip}-2\,\sqrt {2}&-
 \sqrt {2}&0\\\noalign{\medskip}2&0&0\end {array} \right]\,
 \een
etc.
As long as the lack of space does not allow us to
reproduce here the further details,
let us redirect the readers to paper
\cite{[5]} (in which some overall conceptual features of the
EP-related
perturbation approximation construction are described)
and to paper \cite{[12]}
(in which the more complicated EPs with
geometric multiplicity greater than one are taken into consideration).

Out of the most essential conclusions of the latter two
studies let us pick up the single and apparently obvious
fact (still not observed, say, in Refs.~\cite{Trefethen,Viola})
that
the class of admissible, operationally meaningful perturbations
must not violate the self-adjointness of the Hamiltonian in
the correctly reconstructed physical Hilbert space ${\cal H}$.

\subsection{Constructions based on the differential
Schr\"{o}dinger equations.\label{whyxi}}

In the year
1998 Bender with Boettcher
discovered the existence of
the {\em real\,} (i.e., in principle,
experimentally {\em accessible\,}) EPs
generated by certain {\em local\,} and non-Hermitian
but parity-time symmetric (${\cal PT}-$symmetric)
potentials \cite{BB}.
The EPs were
interpreted as instants of
the spontaneous breakdown of ${\cal PT}-$symmetry.
Their reality
was unexpected because
for the conventional local potentials
the EPs are never real \cite{BW}.

Among the specific studies
of the non-Hermitian but
${\cal PT}-$symmetric differential Schr\"{o}dinger
equations
$H\,\psi=E\,\psi$
a distinguished position belongs to paper
\cite{DDT}
by Dorey et al
who considered the
angular-momentum-spiked oscillator
Hamiltonians
 \be
 H(M,L,A)=
 -\,\frac{d^2}{dx^2}
 + \frac{L(L+1)}{x^2} - (ix)^{2M} -A\,(ix)^{M-1}\,,
 \ \ \ \ M=1,2,\ldots\,,
 \ \ \ \  L, A \in \mathbb{R}
 \label{dor}
 \ee
in which the ``coordinate''
$x$ lied on a suitable {\it ad hoc\,}
complex contour. They showed that
inside a suitable domain ${\cal D}$ of parameters
these Hamiltonians generate the strictly real
bound-state-like spectra. These authors
were the first to describe the
shape and role of
the boundaries  $\partial {\cal D}$
formed by the EPs. Unfortunately,
they did not make the picture complete because they did
not construct the corresponding physical inner products.

\subsection{Harmonic oscillator.\label{yxi}}

Once one restricts attention to the most elementary
choice of $M=1$
yielding the one-parametric
harmonic-oscillator Hamiltonian $H(1,L,A)=H^{(HO)}(L)$, the
model becomes exactly solvable at all real $L \in \mathbb{R}$
\cite{ptho}.
For this reason the HO domain of unitarity
${\cal D}^{[HO]}$ has an elementary, multiply connected form of a  ``punched''
interval
with EPs (i.e., with elements $L^{(EP)}$
of boundary $\partial {\cal D}^{[HO]}$)
excluded,
$$
{\cal D}^{[HO]}=
\left(-\frac{1}{2},\frac{1}{2}
\right )
\bigcup
\left(\frac{1}{2},\frac{3}{2}
\right )
\bigcup
\left(\frac{3}{2},\frac{5}{2}
\right )
\bigcup \ldots\,.
$$
\begin{figure}[h]                     %instead of \begin{figure}[t]
\begin{center}                         %instead of \begin{center}
%$\heartsuit$
\epsfig{file=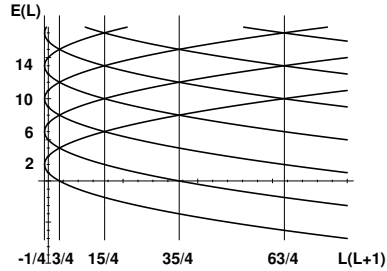,angle=270,width=0.42\textwidth}
\end{center}                         %instead of \end{center}
\vspace{2mm}\caption{Spectrum of Hamiltonian (\ref{dor})
at $M=1$.}\label{HOpic}
\end{figure}
This property (cf. Fig.~\ref{HOpic}) enabled us to
pay more attention, in paper
\cite{[11]}, to
one of the key challenges connected with the
theory, viz., to the constructive analysis of the practical
consequences
of the nontriviality and of the ambiguity of the
related
angular-momentum-dependent metrics $\Theta=\Theta(L)$.
Our main result was
the construction of
a complete menu of the infinite-parametric
assignments $H \to \Theta(H)$
of an eligible metric to the Hamiltonian.

The very possibility
of doing so makes the HO
model truly unique.
For technical as well as phenomenological reasons
we restricted our attention just to the
parameters $L$ which lied close to the points of the
boundary
of the domain of the unitarity, i.e., not far from the set
of EPs
 \be
 \partial {\cal D}= \left \{
 -\frac{1}{2}\,,\
 \frac{1}{2}\,,\
 \frac{3}{2}\,,\ \ \ldots
 \right \}\,.
 \ee
The basic technical ingredient in the construction
of the metrics (see its details
as well as the rather long explicit formulae in \cite{[11]})
was twofold. Firstly, the availability of the
closed-form diagonalization of $H^{(HO)}(L)$
enabled us to replace the
Hamiltonian, at any one of its EP limits, by an equivalent
matrix
called canonical or
Jordan-block representation. Thus, at $L^{(EP)}=-1/2$,
for example, such a representation has the elementary
block-diagonal form
 $$
 %\left [ Q^{{(0)}}_{}
% \right ]^{-1} H^{{(\alpha)}}_{}\,
%  Q^{{(0)}}_{}=
 {J}^{(-1/2)}_{(EP)}=
 \left(
 \begin{array}{cc|cc|cc}
 2&1&0&0&0&\ldots\\
 0&2&0&0&0&\ldots\\
 \hline
 0&0&6&1&0&\ldots\\
 0&0&0&6&0&\ldots\\
 \hline
 0&0&0&0&10&\ldots\\
 \vdots&\vdots&\vdots&\vdots&\ddots&\ddots
 \ea
 \right )\ +\ {\rm corrections}\,.
 $$
Secondly, the highly nontrivial fact that all
of the unavoided
energy-level crossings occurred pairwise and
simultaneously
led to the decomposition of the metric-determining relation
$H^\dagger(L) \Theta(L)=\Theta(L)\,H(L)$ (cf. Eq.~(\ref{cryher}))
to a set of its finite-dimensional (in fact, two-by-two)
matrix components numbered by the separate degenerate
energies $E^{(EP)}=2, 6, 10, \ldots\ $.

In such a setup, every value $L^{(EP)}=-1/2,1/2,3/2,\ldots$ may be
perceived as an instant of a quantum
phase transition which involves all levels
at once.
In a way
accounting, in an exhaustive manner, for the non-uniqueness,
the one-parametric ambiguity
of every two-by-two submatrix of $\Theta(L)$
(once more, recall Eq.~(\ref{secuf}) for illustration)
contributes, independently, to the ultimate
infinite-parametric ambiguity of
selection of the physics-determining inner product in the
infinite-dimensional
physical Hilbert space ${\cal H}^{(HO)}$.

\section{Summary.\label{sekcesedm}}

At present
the Dyson's traditional 3HS recipe
(\ref{picee3wwww})
based on the de-Hermitization interpretation
$\mathfrak{h}\to H $
of Eq.~(\ref{tri})
is usually inverted
to yield the flowchart
 \ben
  %\vspace{-1cm}
  \ba
   %\vspace{-0.3cm}\\
    \begin{array}{|c|}
 \hline
 \vspace{-0.35cm}\\
   {\rm input\!:}\\
   % {\rm one\ picks\ up\ an\ {unphysical} \ but}\\
     {\rm   non\!-\!Hermitian\  } H\
     {\rm with\ real\ spectrum,}\\
  {\rm  \fbox{\rm {\underline{ user-friendly } Hilbert space  {${\cal K}$}}}}\\
   % {\rm {\it i.e.,}\  a\ \bf {\it bona\ fide} \  Hamiltonian};
% \\
  %\vspace{-0.3cm}\\hermitization
  \hline
 \ea
 \stackrel{ {\bf  }  }{ \longrightarrow }
 \begin{array}{|c|}
 \hline
 \vspace{-0.35cm}\\
   {\rm output\!:}\\
   %{\rm  one \ constructs\ an\ eligible } \  {\rm metric\ } \\
   {\rm  metric \  }\Theta=\Omega^\dagger\Omega \ \ ({\rm s.\ t. }\
   H^\dagger \Theta=\Theta
  H),\\
  {\rm  \fbox{\rm {\underline{ physical } Hilbert space {${\cal H}$}}}}\
  % {\rm in \ which }
   \\
  % {\rm the\ evolution\ becomes\ \bf unitary;} \\
   %
%  {\rm {is\ made\   selfadjoint\ {\rm via }}}
% \\
 \hline
 \ea\\
 \ \ \ \ \ \ \ \
%\stackrel{{ \bf the\ Dyson's\ correspondence} }{}
%  \ \nwarrow\!\!\!\searrow\ \  \  \ \ \ \ \ \
 \ \ \ \ \ \ \ \
 \ \ \ \ \ \ \ \
 \ \ \ \ \ \ \ \
 \ \ \ \ \ \ \ \
 \ \ \ \ \  \nearrow\!\!\!\swarrow\
 \stackrel{\bf  equivalent\ predictions}{}
 \\
 \begin{array}{|c|}
  \hline
 \vspace{-0.35cm}\\
   {\rm reference}\!:\\
  %{\rm {for\ a}\ conventional \ {\rm physical}\ interpretation} \\
%  {\rm  one \ may\  reconstruct\  the\  textbook}\\
 {\rm  Hamiltonian }
 \
 \mathfrak{h}
 =\Omega H\Omega^{-1}=
 \mathfrak{h}^\dagger\ \\
  {\rm  \fbox{\rm {\underline{ inaccessible } Hilbert space  {${\cal L}$}}}}\\
 \hline
 \ea \,.
 \\
 \ea
 \label{humandiag}
 \een
The model-building process
is initiated by the choice of a {\it bona fide}
Hamiltonian $H$
which is defined and non-Hermitian
in auxiliary space ${\cal K}$.
The theory is then based on an
exact or approximate
re-Hermitization of $H$ via $\Theta$,
with a very rare or marginal explicit
subsequent reference
to the lower-case Hamiltonian
$\mathfrak{h}$ or to the map of Eq.~(\ref{tri}).
Finally, the variability of parameters in $H=H(\lambda)$
is taken into account, and the physical domain ${\cal D}$
of the admissible values of these parameters
is determined.

In applications
the 3HS formalism is to be kept user-friendly, with
reasonably calculable
predictions.
Besides
the expected enhancement of technical friendliness,
an equally important merit
of the 3HS formalism
should be seen in an emerging access to
new and unusual phenomena.
By our present selection, all of the
phenomena under consideration were characterized by
the proximity of EPs, treated as
forming the boundary $\partial {\cal D}$
of the domains of
``acceptable'' {\it alias\,} ``physical''
(i.e., unitarity-compatible)
parameters of the model in question.
We reviewed and slightly extended
several recent related results.

From the abstract methodical point of view we put
emphasis upon the suitability and
amendments of the necessary
(although, sometimes, less usual) perturbation-type
construction techniques.
This enabled us to clarify several
counterintuitive facts characterizing the
behavior of the closed quantum systems
in the small vicinities of EPs of higher orders.
As a main conclusion the readers should remember
the fact that in these vicinities,
the technique of perturbations offered one of the most efficient tools
of the parametrization and classification of the ``admissible''
(i.e., unitarity-preserving)
multiparametric Hamiltonians
$H(\lambda)=H[a(\lambda), b(\lambda), \ldots, z(\lambda)]$.

\newpage
%.\vspace{-2cm}

%\section*{References}

\end{document}